# Quantum state depression in semiconductor quantum well

Avto Tavkhelidze and Vasiko Svanidze

*Tbilisi State University, Chavchavadze Avenue 13, 0179 Tbilisi, Georgia*

*avtotav@gmail.com*

In this study, the quantum state depression (QSD) in semiconductor quantum well (QW) is investigated. The QSD emerge from the ridged geometry of the QW boundary. Ridges impose additional boundary conditions on the electron wave function and some quantum states become forbidden. State density reduces in all energy bands, including conduction band (CB). Hence, electrons, rejected from the filled bands, must occupy quantum states in the empty bands due to Pauli Exclusion principle. Both the electron concentration in CB and Fermi energy increases as in the case of donor doping. Since quantum state density is reduced, the ridged quantum well (RQW) exhibits quantum properties at widths approaching 200 nm. Wide RQW can be used to improve photon confinement in QW-based optoelectronics devices. Reduction in the state density increases the carrier mobility and makes the ballistic transport regime more pronounced in the semiconductor QW devices. Furthermore, the QSD doping does not introduce scattering centers and can be used for power electronics.

## 1. Introduction

Quantum well (QW) lasers, solar cells, and transistors are fabricated based on semiconductor heterostructure technologies.[1] Typical thickness of the QW layer is 10–20 nm. Lower thickness is essential to reduce the density of the quantum states and realize the quantum properties of the well. However, thin layers do not confine the photons (needed for optoelectronics) and do not carry high currents (needed for power electronics). Recently, quantum state depression (QSD) was investigated both theoretically[2] and experimentally.[3] The QSD allows reduction of quantum state density and realization of quantum properties of the thick layer. It is based on the ridged geometry of the layer boundary. Periodic ridges impose additional boundary conditions on the electron-wave function. Supplementary boundary conditions forbid some quantum states for free electrons, and the state density in $k$ space $\rho(k)$ reduces. Due to the Pauli Exclusion principle, electrons rejected from the forbidden quantum states have to occupy the states with higher $k$. Thus, Fermi vector $k_F$ and Fermi energy $E_F$ increase. In semiconductors, QSD reduces $\rho(E)$ in all energy bands including the conduction band (CB). Electrons, rejected from the filled bands, occupy the quantum states in the empty bands, and the electron concentration in CB increases. This corresponds to donor doping. The QSD depends on electron confinement and, therefore, is most pronounced in QW structures.

The objective of this work is to calculate the parameters of semiconductor ridged quantum well (RQW) and discuss the possible applications. Initially, the basic features of QSD in semiconductors are described. Subsequently, the density of the quantum states in *k* space $\rho(k)$ and energy $\rho(E)$ for RQW is evaluated and compared with those of the conventional QW. The number of QSD generated electrons is determined and the formulas for electron concentration and $E_F$ are obtained. Furthermore, the charge transport in RQW is analyzed and the formulas for carrier mobility and electrical conductivity are obtained. Finally, the advantages of using RQW for optoelectronic devices and ballistic transport devices are discussed.

## 2. Semiconductor RQW

Cross-section of an RQW layer is shown in Fig. 1(a). There are periodic ridges having width *w* and height *a* on the surface of a conventional QW layer. Potential energy changes instantly, by value *D*, at the surface of all walls. Fig. 1(b) shows the schematic representation of the corresponding potential well. Dashed line depicts the double right-side boundary. In addition, the metal RQW was investigated in an earlier study.[2] Let us now find the distinctive features of the semiconductor RQW. Like in metals, the QSD also forbids some quantum states. However, before going into their details, the distinctions and similarities between the QSD forbidden state and a hole should be elucidated. The QSD forbidden state is forbidden by the boundary conditions and cannot be occupied. However, it is not forbidden in an irrevocable way. If the boundary conditions change (e.g., due to charge depletion), then the QSD forbidden state can recombine with the electron. As the QSD forbidden state is confined to the boundary conditions (macroscopic geometry), it is not localized in the lattice and cannot move like a hole.

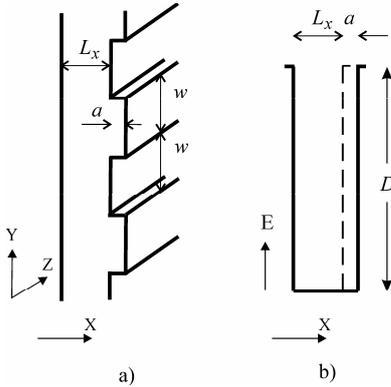

Fig. 1. a) The 3D view of RQW, b) schematic representation of RQW.

The QSD transfers electrons to higher energy levels. If initially the semiconductor is of *p* type, then the QSD will change it to an undoped or even to *n* type. The QSD is comparable with a conventional donor doping, from the point of increase in $E_F$. However, there are no donor atoms in the case of QSD doping, which makes it akin to modulation doping. Unlike modulation doping, there is no space charge, as the QSD does not redistribute the charge and just transfers the electrons from the filled energy bands to the empty ones. Moreover, the material remains uniformly neutral.

It is convenient to make comparison between the RQW and QW. Furthermore, the main parameters, such as $\rho(k)$, $\rho(E)$, and $E_F$, of the RQW can be expressed in terms of the same parameters of conventional QW (*a*=0). It can be assumed that both the wells are made from the undoped material and are deep enough (to allow the limit of infinitely deep well). The $\rho(k)$ is inversely proportional to the volume of *k* space elementary cell. Cell volume for the RQW can be found [2] on the basis of volume perturbation method of solving the time-independent Schrödinger equation (Helmholtz equation).[4, 5] However, this method can only be used when $a<<L_x$. The RQW volume is divided into two parts: main volume (MV) and additional volume (AV). It is presumed that MV>>AV and it defines the form of the solutions for the whole RQW volume. Subsequently, the solutions of the RQW volume are searched in the form of solutions of the MV. The method is especially effective in the case when MV has a simple geometry, e.g., rectangular geometry, allowing separation of the variables. In this study, the volume of the ridge was regarded as AV having dimensions *a, w, $L_z$*. The MV had the dimensions, $L_x, L_y, L_z$. Solutions were plane de Broglie waves with discrete *k* spectrum. Further, the electron-wave function and its derivative were matched from the two sides on the border of MV and AV. The result obtained was the reduction of $\rho(k)$ and the increase of $E_F$ in RQW (detailed description can be found in Ref. 2). Analysis was made within the limit of the quantum model of free electrons. Here, we extrapolate the results to Bloch waves with the assumption that, the electron energy is $E(k) = \hbar^2 k^2 / 2m^*$ and $m^*$ is energy independent, where $\hbar$ is the Planks constant and $m^*$ is an electron effective mass.

The $k$ space elementary cell volumes for RQW and QW are $(2\pi)^3 / awL_z$ and $(2\pi)^3 / L_x L_y L_z$, respectively (as found in Ref. 2). Here, $L_x$, $L_y$, and $L_z$ are the well dimensions. Thus, the corresponding (not normalized) state densities, $P(k)$ are

$$P_{RQW}(k) = \frac{2awL_z}{(2\pi)^3}, \quad P_{QW}(k) = \frac{2L_x L_y L_z}{(2\pi)^3}. \quad (1)$$

Factor 2 accounts for the spin. Thus, the normalized state densities are

$$\rho_{RQW}(k) = \frac{2}{(2\pi)^3} \frac{aw}{L_y(L_x + a/2)}, \quad \rho_{QW}(k) = \frac{2}{(2\pi)^3}. \quad (2)$$

In Eq. (2), the real space volumes of RQW and QW are considered, and we introduced the geometry factor

$$G = L_y(L_x + a/2)/aw. \quad (3)$$

Thus, the comparison in Eq. (2) gives that $\rho_{RQW}(k) = \rho_{QW}(k)/G$. State density in $k$ is reduced by factor $G$. The periodic lattice potential does not depend on QSD, and hence, $m^*$ and the dispersion relation $E(k)$ are identical for both RQW and QW. The state density in energy $\rho(E) = (dE/dk)^{-1} \rho(k)$ is reduced by the same factor $G$, i.e.,

$$\rho_{RQW}(E) = \rho_{QW}(E)/G. \quad (4)$$

Subsequently, the concentration of the QSD-generated electrons $n_{QSD}$ was determined. The quantum well layer was typically grown on a substrate of diverse band structures. A general case shown in Fig. 2 demonstrates that the bandgaps of the substrate material are wider. The QSD takes place within the electron confinement intervals $\Delta E_j$, where j=1, ..., 4. Each $\Delta E_j$ has the characteristic dispersion relation $E_j(k) = \hbar^2 k^2 / 2m_j^*$ and density of states $\rho^{(j)}(E)$. Here, $m_j^*$ is the electron effective mass within the $j$th interval. Inside each $\Delta E_j$, there exist QSD forbidden states, whose densities are

$$\rho^{(j)}_{FOR}(E) = \rho^{(j)}_{QW}(E) - \rho^{(j)}_{RQW}(E) = \rho^{(j)}_{QW}(E)(1 - G^{-1}) \quad (5)$$

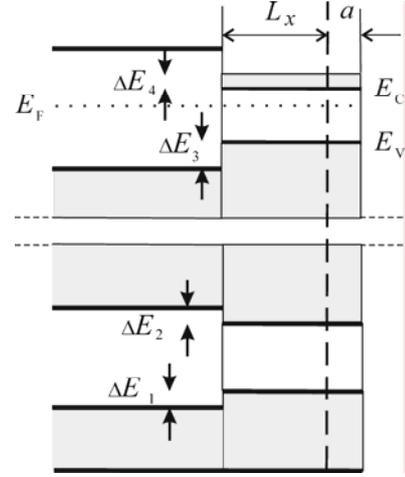

Fig. 2. Energy diagram of semiconductor RQW grown on wide bandgap substrate.

and Eq. (4) was used in Eq. (5). The total density of the forbidden states is the sum of the densities of the forbidden states from all intervals $\Delta E_j$, i.e.,

$$\rho_{FOR} = \sum_j \rho^{(j)}_{FOR}(E) = (1 - G^{-1}) \sum_j \rho^{(j)}_{QW}(E). \quad (6)$$

The sum $\sum_j \rho^{(j)}_{QW}(E)$ depends on the band structures of both the substrate and RQW material, and can be calculated for a particular pair. Apparently, the sum does not depend on the QSD. Thus, the expression $n_{CON} \equiv \sum_j \rho^{(j)}_{QW}(E)$ is introduced, and the index shows that it is the electron confinement-defined number. The $n_{CON}$ does not depend on the energy, since the summation by energy was already carried out. Thus, according to Eq. (6), the total number of forbidden quantum states (per unit volume) or concentration of QSD-generated electrons can be rewritten as

$$n_{QSD} = \rho_{FOR} = n_{CON}(1 - G^{-1}). \quad (7)$$

Equation (7) gives the QSD doping. To calculate other RQW parameters, we used the condition of electrical neutrality.[6]

$$n_{RQW} = p_{RQW} + n_{QSD} = p_{RQW} + n_{CON}(1 - G^{-1}). \quad (8)$$

where $n_{RQW}$ and $p_{RQW}$ are the electron and hole concentrations in the RQW. They can be found using the semiconductor equation for the non-degenerate limit, as follows:

$$n_{RQW} p_{RQW} = \frac{N_C}{G} \frac{N_V}{G} \exp(-E_g / K_B T) = \frac{n_{QW}^2}{G^2} = n_i^2. \quad (9)$$

Here, $N_C$ and $N_V$ are the effective state densities in CB and VB of QW, $E_g = E_c - E_v$ is the bandgap width, $K_B$ is Boltzmann's constant, $T$ is the absolute temperature, $n_{QW}$ is the electron concentration in QW, and $n_i$ is the initial (to QSD doping) electron concentration in RQW. To obtain Eq. (9), we divided the state densities by a factor $G$ according to Eq. (4), and used the semiconductor equation for the conventional QW $n_{QW}^2 = N_C N_V \exp(-E_g / K_B T)$. The combination of Eq. (8) and Eq. (9) gives

$$n_{RQW} = \frac{1}{2G}\left\langle n_{CON}(G-1) + \left[n_{CON}^2(G-1)^2 + 4n_{QW}^2\right]^{1/2}\right\rangle, \quad (10)$$

where $n_{RQW}$ is similar to $n_{QW}$ in the limits of QSD absence. The first limit is $G=1$ (no state density reduction). Equation (10) shows that for $G=1$, $n_{RQW} = n_{QW}$. Another limit is $n_{CON}=0$ (no confinement), in which Eq. (10) gives $n_{RQW} = n_{QW}/G$, where the latter is not similar to $n_{QW}$ for any value of $G$. Divergence is apparent, since $G$ can have only one value of $G=1$ in the case of zero confinement. Actually, $n_{CON}=0$ corresponds to no boundaries and no RQW geometry. The $p_{RQW}$ can be obtained from Eq. (10) and Eq. (9).

Subsequently, we determined the increase in $E_F$ due to QSD doping $\Delta E_F$. We used the formula $\Delta E_F = k_B T \ln(n_{RQW}/n_i)$ for the non-degenerate limit.[6] By inserting $n_{RQW}$ from Eq. (10) and $n_i$ from Eq. (9), we get

$$\Delta E_F = k_B T \ln\left\langle \frac{n_{CON}(G-1)}{2n_{QW}} + \left[\frac{n_{CON}^2(G-1)^2}{4n_{QW}^2} + 1\right]^{1/2}\right\rangle. \quad (11)$$

Figure 3 demonstrates the $\Delta E_F$ dependence on $n_{CON}$ for some values of $G$ according to Eq. (11). The $\Delta E_F$ is most sensitive to changes in $G$ for low $G$ value ($G \approx 1$). For $G>3$, the dependence is less sensitive to changes in $G$ and is linear in the logarithmic scale. Such behavior is natural, as for $G$ with somewhat exceeding unity, the state density remains high and small increase in the value of $G$ generates large number of QSD-rejected electrons. On the contrary, for $G>3$, the state density is reduced dramatically, and further increase of $G$, does not generate that much rejected electrons. In addition, the $\Delta E_F$ will further increase for $(n_{CON}/n_{QW}) > 10^5$. However, we do not extend the curves, since Eq. (11) is true only within the non-degenerate limit. In the case of higher $n_{CON}/n_{QW}$ when the semiconductor becomes degenerated, Fermi integrals should be used to calculate the $n_{RQW}$ and $\Delta E_F$. However, this could be done only within the limited energy range near the bottom of CB, since the above analysis is true only in the approximation that $m^*$ is energy independent.

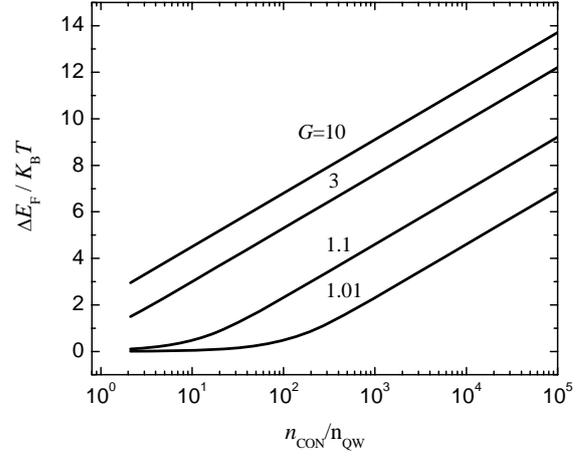

Fig. 3. Fermi energy increase as the function of QSD doping for some values of $G$.

Charge carrier scattering rates are proportional to $\rho(E)$ according to Fermi's golden rule and are reduced in RQW. If $\tau$ is the carrier transport lifetime, then according to Eq. (4), we have $\tau_{RQW} = G\tau_{QW}$. Consequently, for mobility $\mu = e\tau/m^*$, we get

$$\mu_{RQW} = G\mu_{QW}. \quad (12)$$

The mobility of charge carriers increases $G$ times in the RQW. In the case of heavy QSD doping ($n_{RQW} \gg p_{RWQ}$), the hole current can be neglected and the electrical conductivity using Eq. (12) in Eq. (13) is given as

$$\sigma_{RQW} = e\mu_{RQW}\, n_{RQW} = \sigma_{QW}\, G\, (n_{RQW}/n_{QW}). \quad (13)$$

Furthermore, by inserting Eq. (10) in Eq. (13), we get

$$\sigma_{RQW} = \sigma_{QW}\left\langle \frac{n_{CON}(G-1)}{2n_{QW}} + \left[\frac{n_{CON}^2(G-1)^2}{4n_{QW}^2} + 1\right]^{1/2}\right\rangle. \quad (14)$$

Equation (14) indicates that the conductivity of RQW increases with respect to QW. Figure 4 shows the conductivity dependence on QSD-generated electron concentration for some values of $G$. In the general case, the hole current contributes to the mechanism and should be included in Eq. (13), and can be calculated in a similar way using $p_{RQW}$ determined from Eqs (10) and (9).

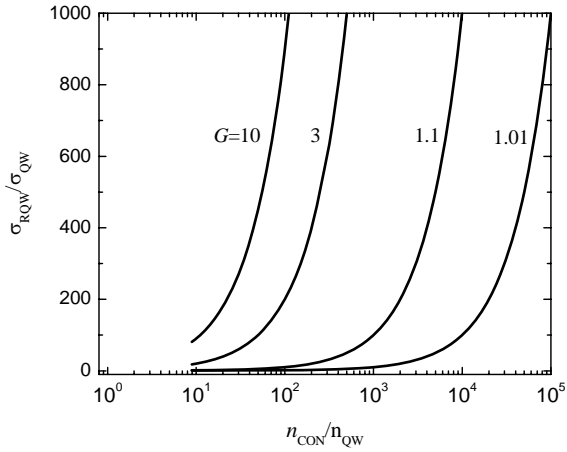

Fig.4. Conductivity dependence on QSD doping for some values of $G$.

## 3. Geometry Factor Calculation

Conventional QW ($a=0$) has quasi-2D structure $L_y, L_z \gg L_x$ and no quantum features are considered in the $Y$ and $Z$ directions. Non-normalized state density is $P(E) = \rho(E) L_x S$, where $S$ is a layer surface, $P(E)$ is proportional to the product $\rho(E) L_x$. As $\rho(E)$ in RQW is reduced $G$ times, $L_x$ can be similarly increased with respect to $G$ times without the loss of quantum properties. Let us find $G$ for the arbitrary geometry. There are no analytical solutions to the time-independent Schrödinger equation in the ridged well (solution contains infinite sums). However, there are fairly accurate mathematical and numerical methods. Mathematically, there is no difference between QSD and electromagnetic mode depression, and Helmholtz equation and the same boundary conditions are used in both the cases. Helmholtz spectrum calculation can be found in the literature related to Casimir effect. Casimir energy exhibits strong dependence on photon spectrum and consequently, on the geometry of the vacuum gap.[7] A number of geometries, including double-side ridged geometry [8] and double-side corrugated geometry [9] were analyzed. New, optical approach for arbitrary geometry was also developed [10]. In addition, a software designed for wave-mode calculation in ridged waveguides has been developed,[11, 12] and can be used to determine $G$ numerically.

In practice, $w \gg a$, which allows the assumption that $k$ spectrum is quasi-continuous in $Y$ direction. Thus, in the first approximation, $G$ can be rewritten in a simpler form as

$$G = (L_x + a/2)/a \approx L_x/a. \quad (15)$$

In Eq. (15), we consider that $a \ll 2L_x$, which is satisfied automatically within the perturbation method limit [Eq. (3) is obtained using perturbation method]. We presume that this method is precise enough in the range of $5<(L_x/a)<10$, and Eq. (15) can be used for that range (the method cannot be used for $a \to 0$, since diffraction leads to ignoring the ridges by wave). Therefore, we used the values of $G= 5 \div 10$ for further estimations. In practice, 20-nm wide conventional QW can be replaced by $100 \div 200$-nm wide RQW.

## 4. Possible Applications

For optoelectronics and power electronics, it is important to have suitable wide-bandgap materials.[13] Semiconductors with $E_g>1.5$ eV are difficult to employ, as their electrical conductivity is low. Hence, donor doping is typically used to increase the conductivity. However, conventional doping introduces impurity centers and increases electron scattering, and thus, the QSD doping can be used to solve these problems. It increases the electron concentration in CB without introducing scattering centers. Besides, the QSD also increases the carrier mobility in both CB and VB.

The QW embedded in *p-i-n* junction is frequently used for solar cells,[14, 15] semiconductor lasers,[16] and infrared detectors.[17] Typical QW layer is only 10–20-nm thick, and thus, there exists the problem of light confinement. To overcome this, complicated multiple QW heterostructures are fabricated. The QSD can contribute in difficulty solving. The RQW layer has the same quantum properties at $G$ times more thickness, and increases the light confinement. Hence, reduced number of RQW layers will be required. In addition, the combination of QW and RQW can also be used for solar cells.

Ballistic MOSFET are widely discussed in the literature.[18, 19, 20] Ballistic regime is difficult to realize in practice, because of low mean free-path (5–10 nm) of charge carriers. Using RQW in the transistor channel reduces the scattering rates, and consequently increases the mean free-path to $G$ times, for both electrons as well as holes.

Molecular Beam Epitaxy (MBE) is typically used to grow quantum well layers. The RQW growth does not differ from a conventional QW growth,

except that the RQW layer has more thickness. Thus, it becomes simpler to fabricate from the point of thickness accuracy. Since RQW is hundreds of nanometers thick, different fabrication methods can also be used, e.g., silicon on insulator (SOI) technology can be utilized to cleave and bond layers of that thickness.[21] The SOI allows mechanical attachment of RQW layer to the substrate, instead of growing it using complicated MBE technology.

## 5. Conclusions

The QSD in the semiconductor RQW was studied. This study demonstrated that QSD reduces the density of quantum states by geometry factor *G,* and the electrons from the filled energy bands are transferred to the empty ones. The electron concentration in CB increases and corresponds to the donor doping. The QSD doping does not introduce impurities, but increases the carrier mobility to *G* times.

Formulas for carrier concentrations and $E_F$ were obtained in the non-degenerate limit. It was observed that the methods developed for Casimir energy calculation in complicated geometries can be utilized to obtain the precise value of *G.*

The RQW exhibited quantum properties at *G-*times more width with respect to the conventional QW. This can be used in various applications—in optoelectronics, the RWQ can increase the light confinement and reduce the number of layers; and in MOS transistors, the RQW can improve the ballistic properties.

**Acknowledgments**

The authors are grateful to A. Bibilashvili for the useful discussions. This work was supported by Borealis Technical Limited, assignee of the corresponding U.S. patents (7,220,984; 7,166,786; 7,074,498; 6,281,514; 6,495,843; 6,680,214; 6,531,703; and 6,117,344).